\documentclass[11pt]{article}
\usepackage{color}

\usepackage{amsmath, amssymb,amsfonts,longtable,tabularx}
\usepackage{cite}
\usepackage{colortbl}
\usepackage[all]{xy}
\usepackage{epsfig}
\usepackage{subfigure}
\usepackage{graphics}

\usepackage{setspace}
\usepackage{caption}
\usepackage{pifont}
\usepackage[table]{xcolor}

\definecolor{gray1}{rgb}{0.9,0.9,0.9}
\definecolor{gray2}{rgb}{0.8,0.8,0.8}

\newcommand{\ZD}{\textit{ZD}}
\newcommand{\good}{\textit{good}}

\usepackage[hmargin=2cm, vmargin=2cm]{geometry}

\date{}

\usepackage{cite}

\begin{document}

\begin{flushleft}
{\huge
\textbf{From extortion to generosity, \\  evolution in the Iterated Prisoner's Dilemma}
}
\bigskip
\\
Alexander J. Stewart$^{1}$, 
Joshua B. Plotkin$^{1,2}$
\\
\bigskip
\bigskip
$^1$ Department of Biology, University of Pennsylvania, Philadelphia, PA 19104, USA
\\
$^2$ E-mail: jplotkin@sas.upenn.edu\\
\end{flushleft}

\vspace{1cm}

%\begin{abstract} 
%\end{abstract}

%\section*{Abstract}

\noindent \textbf{Recent work has revealed a new class of ``zero-determinant"
(\textit{ZD}) strategies for iterated, two-player games.  \textit{ZD} strategies
allow a player to unilaterally enforce a linear relationship between her score and
her opponent's score, and thus achieve an unusual degree of control over both
players' long-term payoffs. Although originally conceived in the context of
classical, two-player game theory, \textit{ZD} strategies also have consequences
in evolving populations of players.  Here we explore the evolutionary prospects
for \textit{ZD} strategies in the Iterated Prisoner's Dilemma (IPD).  Several
recent studies have focused on the evolution of ``extortion strategies'' -- a
subset of zero-determinant strategies -- and found them to be unsuccessful in
populations.  Nevertheless, we identify a different subset of \textit{ZD}
strategies, called ``\textit{generous ZD} strategies," that forgive defecting
opponents, but nonetheless dominate in evolving populations.  For all but the
smallest population sizes, \textit{generous ZD} strategies are not only robust to
being replaced by other strategies, but they also can selectively replace any
non-cooperative \textit{ZD} strategy. \textit{Generous} strategies can be
generalized beyond the space of \textit{ZD} strategies, and they remain robust to
invasion.  When evolution occurs on the full set of all IPD strategies, selection
disproportionately favors these \textit{generous} strategies.  In some regimes,
\textit{generous} strategies outperform even the most successful of the
well-known Iterated Prisoner's Dilemma strategies, including win-stay-lose-shift.}

\section*{Introduction}

\noindent  Press and Dyson \cite{Press:2012fk} recently revealed a remarkable class
of strategies, called ``zero-determinant'' (\textit{ZD}) strategies, for iterated
two-player games.  \textit{ZD} strategies are of particular interest in the
Iterated Prisoner's Dilemma (IPD), the canonical game used to study the emergence
of cooperation among rational individuals
\cite{Rapoport:1965uq,Axelrod:1981kx,Axelrod2,Fund1,Fund2,Nowak:1993vn,Trivers,Boer}.
By allowing a player to unilaterally enforce a linear relationship between her
payoff and her opponent's payoff, Press and Dyson argue, \textit{ZD} strategies
provide a sentient player unprecedented control over the longterm outcome of
Iterated Prisoner's Dilemma games.  In particular, Press and Dyson highlighted a
subset of \textit{ZD} strategies, called ``extortion strategies," that
grant the extorting player a disproportionately high payoff when employed against
a naive opponent who blindly adjusts his strategy to maximize his own payoff.

A natural response to Press and Dyson is to ask: What are the implications of
\textit{ZD} strategies for an evolving population of players
\cite{Stewart:2012ys}?  Although several recent studies have begun to explore this
question \cite{Adami,Sig2}, they have focused almost exclusively on extortion
strategies.  Extortion strategies are not successful in evolving populations,
unless the population size is very small.  Like all strategies that prefer to
defect rather than to cooperate, extortion strategies are vulnerable to strategies
that reward cooperation but punish defection.  However, there is more to
\textit{ZD} strategies than just extortion, and recent work has uncovered some
\textit{ZD} strategies that promote cooperation in two-player games
\cite{Akin,Stewart:2012ys}.  Here we consider the full range of \textit{ZD}
strategies in a population setting and show that, when it comes to evolutionary
success, it is generosity, not extortion, that rules.  

We begin our analysis by considering populations restricted to the space of
\textit{ZD} strategies. We show that evolution within \textit{ZD} always
leads to a special subset of strategies, which we call \textit{generous ZD}.
\textit{Generous ZD} strategies 
%are qualitatively similar to the well-known strategy generous tit-for-tat.  Like
%generous tit-for-tat, generous ZD strategies cooperate with cooperators, and
%mostly punish defectors, but will
reward cooperation, but punish defection only mildly, and they tend to score lower
payoffs than defecting opponents.  Next, we build on recent work by Akin
\cite{Akin}, who identified \textit{generous} strategies beyond those contained
within \textit{ZD}.  We demonstrate that a large proportion of these
\textit{generous} strategies are robust to replacement in an evolving population.
The robust, \textit{generous} strategies can at worst be replaced neutrally.  Conversely, 
we demonstrate that most \textit{generous} strategies can readily replace resident non-\textit{generous} 
strategies in a
population.  As a result, \textit{generous} strategies are just as, or
sometimes even more, successful than the most successful of well-known Iterated
Prisoner's Dilemma strategies in evolving populations.  Finally, we show that
populations evolving on the full set of Iterated Prisoner's Dilemma strategies
spend a disproportionate amount of time near \textit{generous} strategies,
indicating that they are favored by evolution.
%
%Our analysis, which relies inherently on our knowledge of ZD strategies, reveals
%an important new subset of evolutionarily successful strategies, and shows that,
%in evolution, generosity works better than extortion.

\section*{Methods and Results}

In the Prisoner's Dilemma, two players $X$ and $Y$ must  simultaneously
choose whether to cooperate ($c$) or defect ($d$).  If both players cooperate
($cc$) then they each receive payoff $R$.  If $X$ cooperates and $Y$ defects
($cd$), then $X$ loses out and receives the smallest possible payoff, $S$, while
$Y$ receives the largest possible payoff, $T$.  If both players defect ($dd$),
then both players receive payoff $P$.  Payoffs are specified so that the reward
for mutual defection is less than the reward for mutual cooperation,
i.e.~$T>R>P>S$. It is typically assumed that $2R>T+S$, so that it is not
possible for total payoff received by both players to exceed $2R$.
In what follows we will consider the payoffs $T=B$, $R=B-C$, $P=0$ and $S=-C$,
which comprise the so-called ``donation game'' \cite{Sig2}.

The Iterated Prisoner's Dilemma (IPD) consists of infinitely many successive
rounds of the Prisoner's Dilemma.  Press and Dyson \cite{Press:2012fk} showed that
it is sufficient to consider only the space of memory-1 strategies,
i.e.~strategies that specify the probability a player cooperates each round in
terms of the payoff she received in the previous round.  Memory-1 strategies
consist of four probabilities, $\mathbf{p}=\{p_{cc},p_{cd},p_{dc},p_{dd}\}$.  In
particular, Press and Dyson \cite{Press:2012fk} showed that the longterm payoff to
a memory-1 player pitted against an arbitrary opponent is the same as her payoff
would be against some other, memory-1 opponent. Thus, we limit our analysis to
memory-1 players without loss of generality (see also Materials and Methods).

\subsection*{Evolutionary game theory} In the context of evolutionary game theory,
we consider a population of $N$ individuals each characterized by a strategy
$\mathbf{p}$. We say strategy $\mathbf{p}$ receives longterm payoff
$\pi(\mathbf{p},\mathbf{q})$ against an opponent with strategy $\mathbf{q}$. The
success of a strategy depends on its payoff when pitted against all individuals in
the population \cite{Maynard,Maynard-Smith:1982vn,Hofbauer:1998ys,Boyd:2010zr}.
Traditionally, the evolutionary outcome in such a population has been understood
in terms of evolutionary stable strategies (ESS).  A strategy $\mathbf{p}$ is an
ESS if its longterm payoffs satisfy
$\pi(\mathbf{p},\mathbf{p})>\pi(\mathbf{p},\mathbf{q})$, or
$\pi(\mathbf{p},\mathbf{p})=\pi(\mathbf{p},\mathbf{q})$  and
$\pi(\mathbf{p},\mathbf{q})>\pi(\mathbf{q},\mathbf{q})$, for all opponents
$\mathbf{q}\neq\mathbf{p}$.  

The ESS condition provides a useful notion of stability in the context of an
infinite population.  However, in a finite population, the concept must be
generalized to consider whether selection favors both invasion and replacement of
a resident strategy by a mutant strategy \cite{Nowak:2004fk,Nowak:2006ly}.  In a
finite, homogenous population of size $N$, a newly introduced neutral mutation
(i.e.~a mutation that does not change the payoff to either player) will eventually
replace the entire population with probability $\rho=1/N$.  A deleterious
mutation, which is opposed by selection, will fix with probability $\rho<1/N$;
while an advantageous mutation, which is favored by selection, will fix with
probability $\rho>1/N$.  We say that a resident strategy $\mathbf{p}$ in a finite
population of size $N$ is \textit{evolutionary robust} against a mutant strategy
$\mathbf{q}$ if the probability of replacement satisfies $\rho \leq 1/N$; in other
words, the robust strategy cannot be selectively replaced by the mutant strategy.
In the limit of infinite population size, $N\to\infty$, the condition $\rho<1/N$
reduces to the ESS condition.  

%Thus in a finite population, the ESS concept can be generalized to those
%strategies for which selection opposes replacement by any other strategy
%\cite{Nowak:2006ly}.
When selection is weak ($\sigma \ll 1$), we can write down an explicit criterion
for robustness: a resident $Y$ is evolutionary robust against a mutant $X$ if and
only if (see Materials and Methods and \cite{Nowak:2004fk,Nowak:2006ly})
\begin{equation} s_{xx}(N-2)+s_{xy}(2N-1)\leq s_{yx}(N+1)+2s_{yy}(N-2),
\end{equation}  where we denote the longterm payoff of player $X$ against player
$Y$ by $s_{xy}$.
%and of player $Y$ against player $X$ by $s_{yx}$.  In this paper we identify
%Iterated Prisoner's Dilemma strategies that are evolutionary robust against all
%other strategies.  The evolutionary robustness of a strategy depends on the
%payoffs revived by two players. 
We restrict our analysis to memory-1 players.  In the two-player setting this
restriction does not sacrifice generality, because, as per Press \& Dyson
\cite{Press:2012fk}, the payoff received by a memory-1 strategy $Y$ can be
determined independently of her opponent's memory.  However, in an evolutionary
setting, $Y$'s success depends also on the payoff her opponent receives against
himself.  
%Thus a long-memory player may still gain an evolutionary advantage.  
Nonetheless we will show that our results for \textit{generous} strategies hold
against all opponents, no matter how long their memories, provided the standard
IPD assumption $2R>T+S$ holds. 

\subsection*{Zero-determinant strategies, extortion, and generosity} Among the
space of all memory-1 IPD strategies, Press and Dyson identified a subspace of
so-called ``zero-determinant"  (\textit{ZD}) strategies that ensure a fixed,
linear relationship between two players' longterm payoffs.  If player $Y$ facing
player $X$ employs a \textit{ZD} strategy of the form 
\begin{eqnarray} \nonumber
p_{cc}&=&1-\phi(1-\chi)(B-C-\kappa)\\ \nonumber p_{dc}&=&1-\phi\left[\chi
C+B-(1-\chi)\kappa\right]\\ \nonumber p_{cd}&=&\phi\left[\chi
B+C+(1-\chi)\kappa\right]\\ \nonumber p_{dd}&=&\phi(1-\chi)\kappa, 
\end{eqnarray} 
then their payoffs will satisfy the linear relationship
\begin{equation} 
\label{linear}
\phi\left[s_{xy}-\chi s_{yx}-(1-\chi)\kappa\right]=0.
\end{equation}
The parameters $\chi$ and $\kappa$ must lie in the range
$0\leq\kappa\leq<B-C$ and $\max\left(\frac{\kappa-B}{\kappa+C},\frac{\kappa+C}{\kappa-B}\right)\leq\chi\leq1$
to produce a feasible strategy.  Eq.~2
defines the full space of \textit{ZD} strategies introduced by Press \& Dyson
\cite{Press:2012fk}.  Within this space, two particular subsets are of special
interest: the extortion strategies, described by Press \& Dyson, for which
$\kappa=P=0$ and $\chi>0$, and the \textit{generous} strategies, described in our
commentary \cite{Stewart:2012ys}, for which $\kappa=R=B-C$ and $\chi>0$. 
 
Extortion strategies ensure that either the extortioner $Y$ receives a higher
payoff than her opponent $X$, $s_{yx}>s_{xy}$, or otherwise both players receive
the payoff for mutual defection, $s_{yx}=s_{xy}=0$.  In contrast,
\textit{generous} strategies ensure that both players receive the payoff
for mutual cooperation, $s_{yx}=s_{xy}=B-C$, or otherwise the \textit{generous}
player $Y$ receives a lower payoff than her opponent, $s_{yx}<s_{xy}$.

Recent work has focused on the evolutionary prospects of extortioners
\cite{Adami,Sig2}, and found that such strategies are unsuccessful, except in very
small populations.  In fact, as we will show below, selection favors replacement of
extortioners by \textit{generous} strategies; whereas \textit{generous}
strategies are robust to replacement by extortioners. Moreover, the success of
\textit{generous} strategies persists when evolution proceeds within the full
space IPD strategies.

\subsection*{Evolution of generosity within \textit{\normalsize ZD} strategies} 

We start by identifying the subset of \ZD{} strategies that are evolutionary robust against all IPD strategies
in a population of size $N$.
Substituting Eq.~2 into Eq.~1 shows that a resident \ZD{} strategy $Y$ with $\kappa_y=B-C$ is robust against
any mutant IPD strategy $X$ if and only if $\chi_y \geq (N+1)/(2N-1)$ (see Materials and Methods).
Conversely, provided $N>2$, any resident \ZD{} strategy $Y$ with $\kappa_y<B-C$ can be 
selectively replaced by another
strategy -- namely, by a \ZD{} strategy with $\kappa_x = B-C$ and $\chi_x > (N+1)/(2N-1)$ (see Materials and Methods).
Hence, those \textit{ZD} strategies with $\kappa=B-C$ and
$\chi \geq (N+1)/(2N-1)$ are precisely the \textit{ZD} strategies that are evolutionary robust against all IPD strategies.  We
denote this set of robust \textit{ZD} strategies as \textit{ZD}$_R$: 
%\begin{equation*}
\begin{center}\vspace{0.5cm} 
\textit{ZD}$_R=\left\{(\kappa,\chi,\phi)|\kappa=B-C,1>\chi \geq \frac{N+1}{2N-1}\right\}$.
%\end{equation*} 
\end{center} \vspace{0.5cm} 
Here $\phi$ is left unconstrained, but it must lie in
the range required to produce a feasible strategy, $0<\phi\leq \chi B/(\chi C+B)$.

The robust \textit{ZD} strategies are what we call \textit{cooperative}, meaning
they satisfy $\kappa=B-C$. Any \textit{cooperative} player will 
agree to mutual cooperation when facing another \textit{cooperative} player, and
so they each receive payoff $B-C$.  If
a \textit{cooperative} strategy further satisfies the condition $\chi>0$ then we
say that the strategy is \textit{generous} -- meaning that any deviation from
mutual cooperation causes the \textit{generous} player's payoff to decline more
than that of her opponent. The robust \ZD{} strategies are all \textit{generous}.

We now consider evolution in a population of $N>2$ players restricted to the space of
\textit{ZD} strategies. Because selection favors replacement of any non-\textit{cooperative} \textit{ZD} strategy 
by some member of \textit{ZD}$_R$, we expect evolution
within the space of \textit{ZD} strategies to tend towards \textit{generous}
strategies -- and thereafter to remain at \textit{generous} strategies, because \textit{ZD}$_R$ is robust.
This expectation is confirmed by Monte-Carlo simulations of 
well-mixed populations of IPD players (Fig.~1).  Following
\cite{Sig2,Traulsen:2006zr}, we modeled evolution as a process in which
individuals copy successful strategies with a probability that depends on their
relative payoffs (see Materials and Methods). As Fig.~1 shows, evolution within the set of
\textit{ZD} strategies proceeds from extortion ($\kappa=0$ and $\chi>0$) to
generosity ($\kappa=B-C$ and $\chi>0$).  In fact, even populations initiated with
$\chi<0$ evolve to generosity (Fig.~S1). 

\begin{figure}[h!] \centering \includegraphics[scale=0.5]{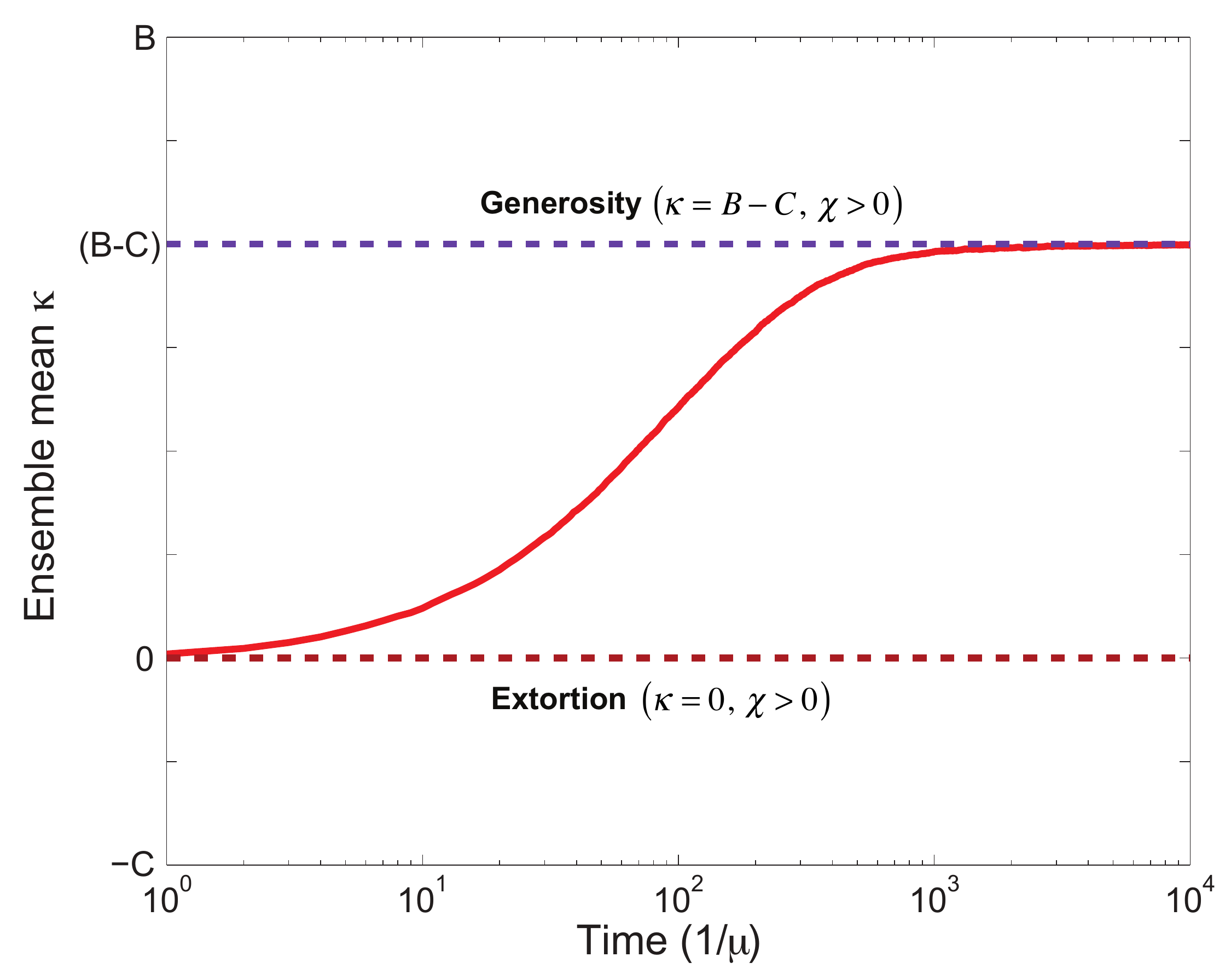}
\caption{\small Evolution from extortion to generosity within the space of
\textit{ZD} strategies. Populations were simulated in the regime of weak mutation. The
figure shows the ensemble mean value of $\kappa$ in the population, plotted over time.
$\kappa=0$ corresponds to the extortion strategies of Press \& Dyson,
whereas $\kappa=B-C$ corresponds to the \textit{generous ZD} strategies. Each
population was initialised at an extortion strategy $E_\chi$, with $\chi$ drawn
uniformly from the range $\chi\in(0,1]$.  Given a resident strategy in the
population, mutations to $\kappa$ were proposed as normal deviates
of the resident strategy, truncated to constrain $\kappa \in [0,B-C]$, while mutations to $\{\chi, \phi\}$ 
were drawn uniformly from $\max\left(\frac{\kappa-B}{\kappa+C},\frac{\kappa+C}{\kappa-B}\right)\leq\chi\leq1$
with $\phi$ drawn uniformly within the feasible range given $\kappa$ and $\chi$. A proposed mutant
strategy replaces the resident strategy with a fixation probability dependent on 
their respective payoffs, as in \cite{Sig2,Traulsen:2006zr}. The mean $\kappa$
among $10^3$ replicate populations is plotted as a function of time.
Parameters are $B=3$, $C=1$, $N=100$, and selection strength $\sigma=1$.} \end{figure}

\subsection*{\textit{\normalsize Good} strategies} The \textit {generous ZD}
strategies identified above are best understood by comparison with the space of
\textit{good} strategies recently introduced by Akin \cite{Akin}.  A \textit{good}
strategy stabilizes cooperative behavior in the two-player IPD: by defintion, if both players
adopt \textit{good} strategies then each receives payoff $B-C$ and
neither player can gain by unilaterally changing strategy.  All \textit{good}
strategies are \textit{cooperative}, i.e.~they have $\kappa=B-C$.
Moreover, the \textit{generous ZD} strategies described above are precisely the
intersection of \textit{good} strategies \cite{Akin} and \textit{ZD} strategies
\cite{Press:2012fk} (Fig.~2).

\begin{figure}[h!] \centering \includegraphics[scale=0.5]{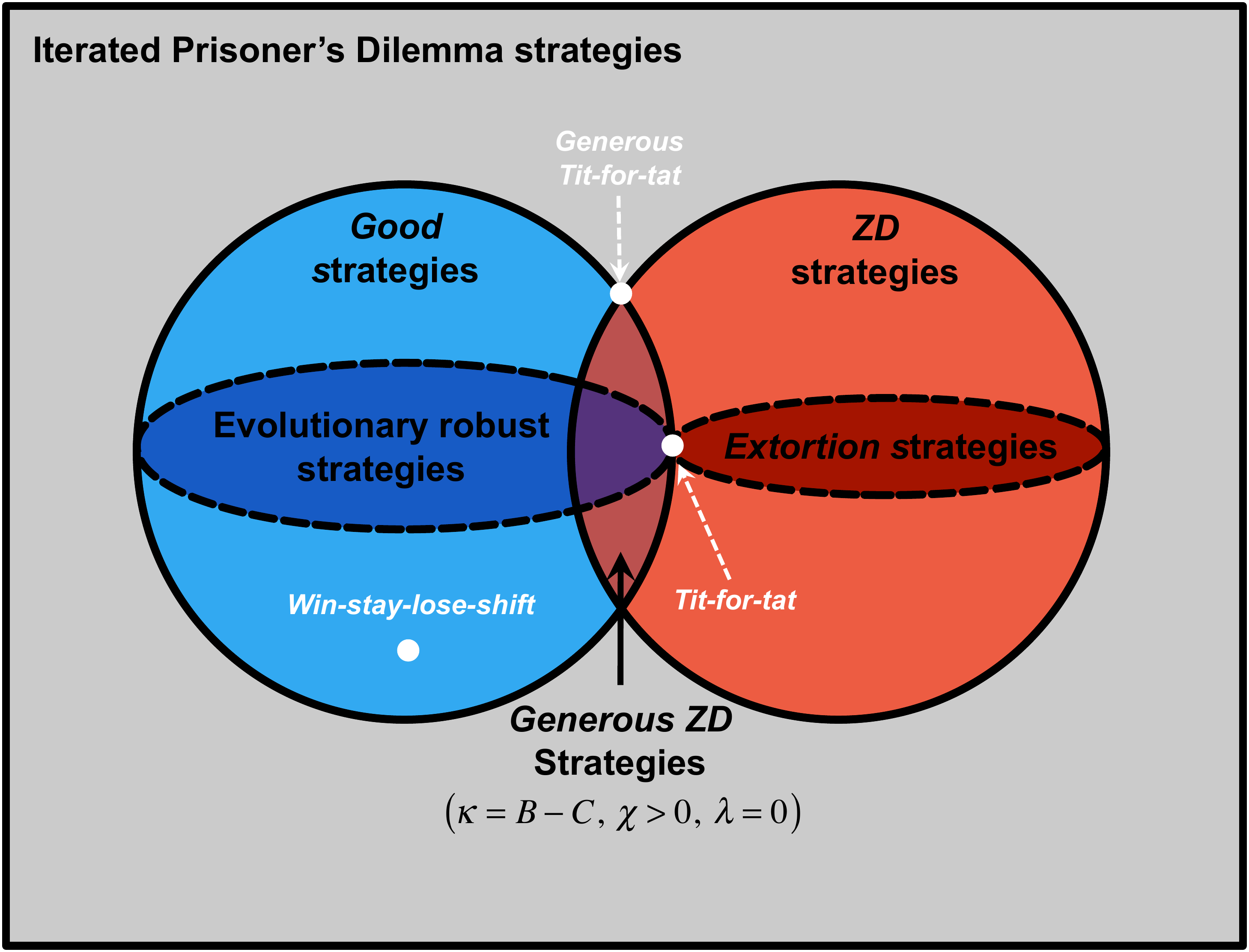}
\vspace*{.2in}
\caption{\small The relationship between \ZD{} and \good{} strategies in the
Iterated Prisoner's Dilemma. The intersection between \ZD{} and \good{} is precisely
the set of \textit{generous ZD} strategies. Not all \good{} strategies are
\textit{generous}. As a result, only a strict subset of \good{} strategies are
evolutionary robust, just as a strict subset of \ZD{} strategies are evolutionary
robust.  Extortion strategies are neither \textit{generous} nor evolutionary robust. Also
shown are the locations of the classical IPD strategies \cite{Nowak:2006ly} win-stay-lose-shift,
tit-for-tat, and generous tit-for-tat. 
} \end{figure}

We can identify the space of memory-1 \textit{good} strategies as those of the form
\begin{eqnarray} \nonumber p_{cc}&=&1-\phi(1-\chi)(B-C-\kappa)\\ \nonumber
p_{dc}&=&1-\phi\left[\chi C+B-(1-\chi)\kappa+\lambda\right]\\ \nonumber
p_{cd}&=&\phi\left[\chi B+C+(1-\chi)\kappa-\lambda\right]\\ \nonumber
p_{dd}&=&\phi(1-\chi)\kappa,
\end{eqnarray}
where $-1\leq\chi\leq1$ and  $-(\chi B+C)\leq\lambda\leq (B+\chi C)$ are required to 
produce a feasible strategy. Sufficient conditions for set $G$ of \good{} strategies 
are (see Supporting Information)
\begin{align}
\nonumber G=&\left\{(\kappa,\chi,\phi,\lambda)| \right. \\
\nonumber &\left. \kappa=B-C, \lambda>-(B-C)\chi,\lambda>-(B+C)\chi\right\},
\end{align}
where the parameter $\phi$ is left unconstrained except that it must produce a feasible
strategy. Numerics indicate these sufficient conditions are also neccesary (see Supporting
Information). Note that the \textit{good} strategies with
$\lambda=0$ correspond precisely to the \textit{generous ZD} strategies.

It is interesting to note that, in addition to tit-for-tat and generous
tit-for-tat, which are \textit{ZD}, the set of \textit{good} strategies also
contains win-lose-stay-shift, which is widely known as one of the most
evolutionary successful IPD strategies \cite{Nowak:1993vn}.  Nonetheless, 
even though win-lose-stay-shift is \textit{good}, it is not \textit{generous} (it has
$\chi=-C/B < 0$, Fig.~3). Because it lacks generosity, win-lose-stay-shift can in fact 
be out-competed in evolving populations, as we shall see below.

\begin{figure}[h!] \centering \includegraphics[scale=0.5]{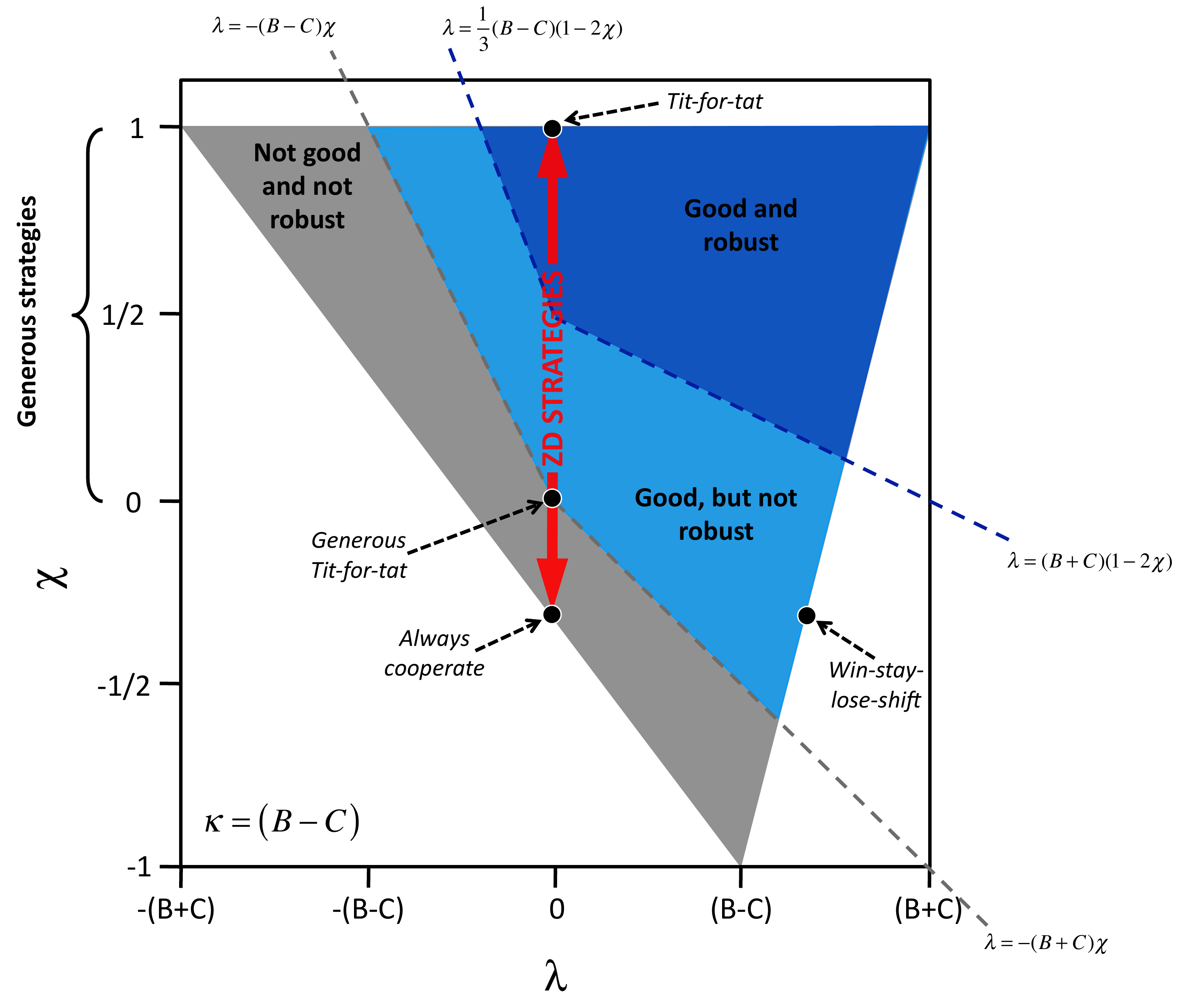}
\caption{\small The space of all \textit{cooperative} IPD strategies, projected onto
the parameters $\chi$ and $\lambda$. The boundary of the simplex delineates the
set of feasible strategies with $\kappa=B-C$. Strategies colored light blue or
dark blue are \good{}; whereas strategies colored dark blue are both \good{} and
evolutionary robust, under weak selection. 
Setting $\lambda=0$ recovers the space of cooperative \ZD{}
strategies (red line).  Note that all robust strategies are \textit{generous}
(i.e.~$\chi>0$ and $\kappa=B-C$).  Each point in the figure, $\{\chi,\lambda\}$,
has an associated range of $\phi$ values, and thus corresponds to multiple IPD
strategies. However, the evolutionary robust \textit{good} strategies resist
replacement by any other strategy, regardless of the choice of $\phi$. 
The figure illustrates the robust region for a large population size;
whereas Eq.~3 gives the exact $N$-dependent conditions for robustness.
Also
shown are the locations of several classical IPD strategies.
Tit-for-tat
($\chi=1$) and generous tit-for-tat ($\chi=0$) are limiting cases of
\textit{generous ZD} strategies, but they are not robust.
Likewise, win-stay-lose-shift is \good{} but not robust.
} \end{figure}

\subsection*{Evolutionary of generosity within \textit{\normalsize good} strategies}

In this section we ask which \good{} strategies are evolutionary robust; and we find
that the robust \good{} strategies are always \textit{generous} -- i.e.~have
$\chi>0$, regardless of $\lambda$.
In the case of \ZD, the conditions for evolutionary robustness do not depend on the
parameter $\phi$. Similarly, we will derive conditions for the robustness of \good{}
strategies that hold regardless of $\phi$.

Application of Eq.~1 allows us to derive the conditions for a \good{}
strategy to be evolutionary robust against all IPD strategies in 
a population of size $N$ (see Supporting Information).  
The resulting set, $G_R$, of evolutionary robust \textit{good} strategies satisfies
\begin{align}
\nonumber G_R=&\Big\{(\kappa,\chi,\phi,\lambda)|\kappa=B-C,\chi<1,\\
\nonumber &\lambda>\frac{B-C}{3N}\left[N+1-(2N-1)\chi\right],\\
&\lambda>\frac{B+C}{N-2}\left[N+1-(2N-1)\chi\right]\Big\}.
\end{align}
Here $\phi$ is left unconstrained except that it must produce a feasible strategy.
These analytical conditions for robustness are confirmed by Monte-Carlo
simulations (Fig.~S2).  Setting $\lambda=0$ in the equation above
recovers the conditions we previously derived for the robustness of \textit{ZD}
strategies. As in the case of \ZD, the robust \textit{good}
strategies are exclusively limited to \textit{generous} strategies (i.e.~strategies
with $\kappa=B-C$ and $\chi>0$, see Fig.~3).  

Interestingly, the strategy win-stay-lose-shift does not lie within the region of
robust \good{} strategies (Fig.~3). As a concrete demonstration of this result we
have identified a specific strategy that selectively replaces
win-lose-stay-shift in a finite population (see Supporting Information and Fig.~S3).
Furthermore, even under strong selection, and under increased mutation rates,
win-stay-lose-shift can be dominated by some strategies (Fig.~S3).

\subsection*{The evolutionary success of generosity}

We have shown that \textit{generous} strategies are evolutionarily robust and
eventually dominate in a population, when players are confined to the space of
\textit{ZD} strategies.  We have also shown that amongst the \textit{good}
strategies, which stabilize cooperative behavior, the evolutionary robust
strategies, $G_R$, are also \textit{generous}.  To complement these results, we
now systematically query the evolutionary success of \textit{generous} strategies
in general, by allowing a population to explore the full set of memory-1
strategies $\mathbf{p}=\{p_{cc},p_{cd},p_{dc},p_{dd}\}$ and quantifying how much
time the population spends near generosity.  

Following \cite{Sig2, Imhof:2010kx}, we performed simulations in the regime of
weak mutation, so that the population is monomorphic for a single strategy at all
times.  Mutant strategies, drawn uniformly from the space
$\{p_{cc},p_{cd},p_{dc},p_{dd}\}$, are proposed at rate $\mu$. A proposed mutant
either immediately fixes or is immediately lost from the population, according to
its fixation probability calculated relative to the current strategy in the
population \cite{Sig2,Traulsen:2006zr}. Over the course of this simulation we
quantified how much time the population spends in a $\delta$-neighborhood of
\textit{ZD}, \textit{ZD}$_R$, \textit{G}, and \textit{G}$_R$ strategies, as well as
extortion strategies (Fig.~4).  The $\delta$-neighborhood of a strategy set is
defined as those strategies within Euclidean distance $\delta$ of it, among the space
of all memory-1 strategies.  If the proportion of time spent in the
$\delta$-neighborhood is greater than would be expected by
random chance (which is proportional to the volume of the $\delta$-neighborhood),
then evolution is said to favor that set of strategies.

It is already known that, except for very small populations, a population spends
far less time near extortion strategies than expected by random chance, and that
the same is true for the set of all \textit{ZD} strategies\cite{Sig2,Adami}. Thus,
extortion and \textit{ZD} strategies in general are disfavored by evolution in
populations. This has led to the view the \textit{ZD} strategies are of importance
only in the setting of classical two-player game theory, and not in evolving
populations \cite{Ball}. However, in Fig.~4 we repeat this analysis but we
additionally report the $\delta$-neighborhoods of \textit{ZD}$_R$, \textit{G}, and
\textit{G}$_R$ strategies.  We find that, except in very small populations,
selection strongly favors \textit{G}, \textit{G}$_R$, and, especially,
\textit{ZD}$_R$ strategies.  In particular, the population spends more than
100-times longer in the neighborhood of \textit{ZD}$_R$ strategies than expected
by random chance. Thus, \ZD{} contains a subset of strategies that are remarkably
successful in evolving populations. 

We also analyzed the time spent near each individual \textit{good} strategy, 
under both weak and strong selection. We found that the strategies most
strongly favored by selection are virtually all \textit{generous} (Fig.~S4). 
The remaining \good{} strategies are typically moderately favored by selection, with the
exception of those near win-stay-lose-shift, which are also strongly favored.

\begin{figure}[h!] \centering \includegraphics[scale=0.5]{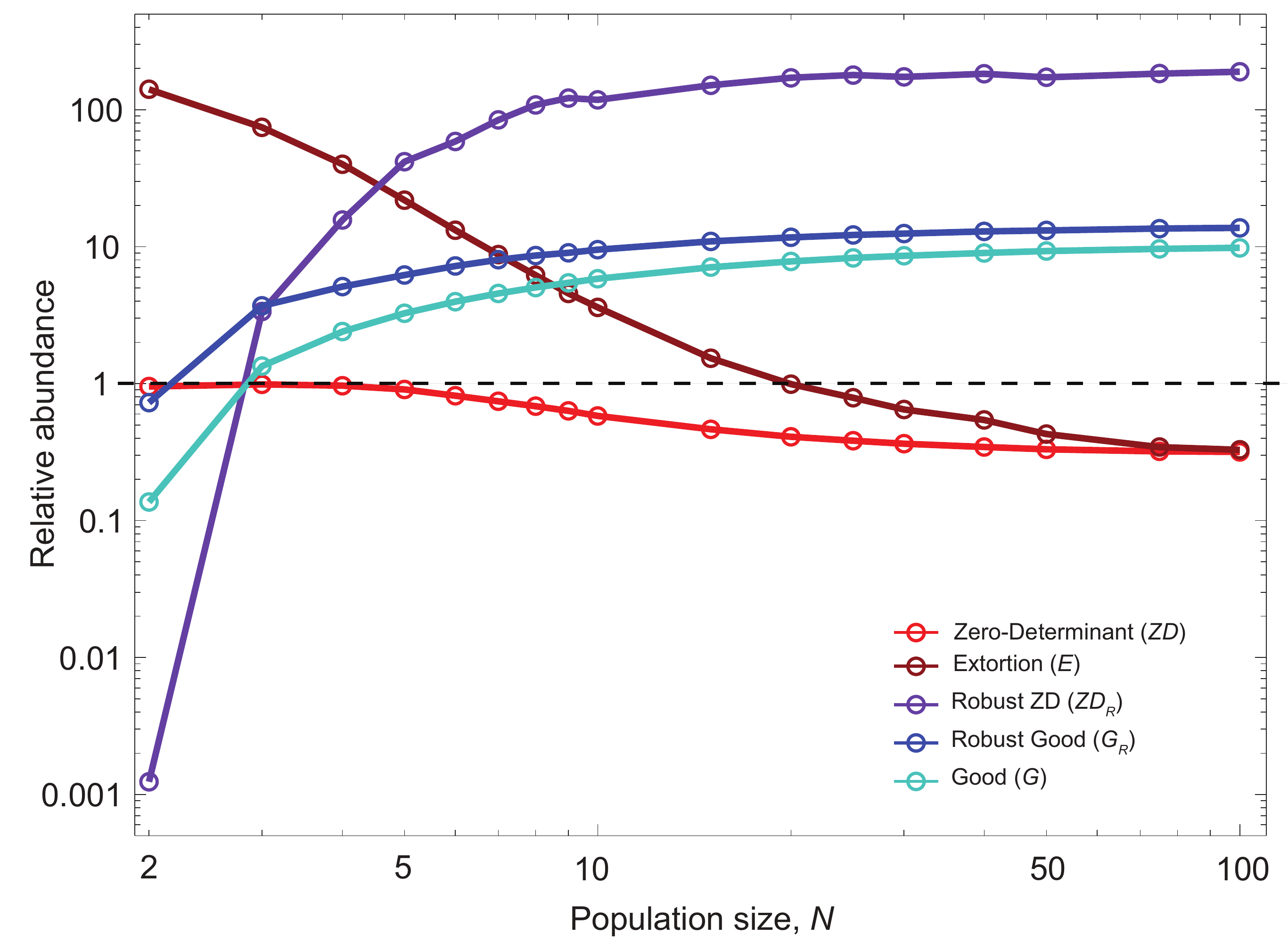}
\caption{\small \textit{Generous} strategies are favored by selection in 
evolving populations. We simulated a population under weak mutation, proposing
mutant strategies drawn uniformly from the full set of memory-1 IPD strategies.  We
calculated the time spent in the $\delta$-neighborhood \cite{Sig2} of \textit{ZD}
and extortion strategies, as well as robust \textit{ZD} strategies, \textit{good}
strategies and robust \textit{good} strategies, relative to their random (neutral)
expectation. For small population sizes, extortioners are abundant and \textit{generous}
strategies are nearly absent. As population size increases, the frequency of
\textit{generous} strategies and \textit{good} strategies is strongly amplified by
selection; whereas extortion strategies, and \textit{ZD} strategies in general,
are disfavored, as previously reported \cite{Sig2}.  Simulations were run until
the population fixed $10^7$ mutations.  Parameters are $B = 3$, $C =1$, $\delta=0.05$, 
and
selection strength $N\sigma=100$.} \end{figure}

\subsection*{The success of \textit{\normalsize generous} strategies against classic IPD strategies}

To complement the weak-mutation studies described above, we also compared the
performance of \textit{generous ZD} strategies against several classic IPD
strategies, in a finite population of players
\cite{Sig2,Traulsen:2006zr,Nowak:2004fk,Nowak:2006ly}, assuming either strong or
weak mutation (i.e.~high or low mutation rates).
%Assuming that there is some ``error rate'', $\epsilon$, in the move played by
%each strategy at each time step \cite{Sigmund:2010ve}, we obtained the payoffs of
%various strategies, in the limit $\epsilon\to0$ (Table S1).  As Table S1 shows,
%\textit{ZD}$_R$ strategies fare very well against most opponents, and so too does
%win-stay-lose-shift, which was previously shown to dominate extortion strategies
%\cite{Sig2}.  As we have seen above, \textit{generous ZD} strategies also easily
%dominates extortion strategies.  
%
%To illustrate the performance of \textit{generous ZD} strategies compared to
%classic IPD strategies, 
We performed Monte-Carlo simulations of populations constrained to different
subsets of strategies, similar to those of \cite{Sig2}.  In these simulations, a
pair of individuals is chosen from the population at each time step, and the first
individual copies the strategy of the second with a probability that depends on
their respective payoffs, as above. Mutations also occur, with probability $\mu$,
so that the mutated individual randomly adopts another strategy from the set of
strategies being considered. We ran simulations at a variety of populations sizes,
ranging from from $N=2$ to $N=1,000$.  

At very small population sizes, defector strategies tend to dominate (Fig.~S5),
reflecting the fact that extortion pays in the classic two-player setting
\cite{Press:2012fk}. However, as the population size increases, \textit{good}
strategies, such as win-stay-lose-shift, and \textit{generous ZD} quickly begin to
dominate (Fig.~S5).  Which strategy does best depends on the population size, the
mutation rate, and the set of available strategies (Figs.~S5-S6).  
In some regimes \textit{generous ZD}
strategies outperform even win-stay-lose-shift (Fig.~S5). 

\section*{Discussion}
We have shown that \textit{generous} strategies tend to dominate in evolving
populations of IPD players. This is a surprising result because, when faced with a defector
strategy, \textit{generous} strategies must by definition suffer a greater
reduction in payoff than their opponent suffers.  And so one might expect
such strategies to be vulnerable to replacement by defector strategies -- whereas
in fact, we have shown, the reverse is true.  Likewise, one might expect
\textit{generous} strategies to be unsuccessful at displacing resident
strategies in a population.  However, simulations reveal (Figs.~S7-S8) that most
\textit{generous} strategies can selectively replace almost all other IPD
strategies.  

How can we account for the remarkable evolutionary success of generosity?  First,
it is important to note that the most successful \textit{generous} strategies are
not \textit{too} \textit{generous}.  For example, in a large population,
evolutionary robust \textit{ZD} strategies must have $\chi>1/2$: that is, they
must reduce their payoff when faced with a defector opponent, but not by too much.
Second, whilst \textit{generous} strategies score less than defector strategies in
head-to-head matches, they are able to limit the difference between their own
payoff and their opponent's payoff (see Materials and Methods).  As a result, they tend to
have a consistent probability $\rho>1/N$ of replacing a diversity of
resident IPD strategies (Fig.~S8), allowing them to succeed in an evolutionary
setting.

We found that \textit{generous ZD} strategies are particularly successful when
mutations arise at an appreciable rate.  Under such circumstances, \textit{ZD}$_R$
strategies can dominate even win-stay-lose-shift, a perennial favourite in
evolving populations \cite{Nowak:1993vn,Imhof:2007uq,Adami,Iliopoulos:2010fk}. Over all,
selection strongly favors \textit{generous} \textit{ZD} strategies when evolution proceeds in
the full space of memory-1 strategies.  These results strongly contravene the view
that \textit{ZD} strategies are of little evolutionary importance \cite{Ball}. In
fact, we have shown that a subset of \textit{ZD} strategies, the \textit{generous}
ones, are strongly favored in the evolutionary setting.

The discovery and elegant definition of \textit{ZD} strategies remains a
remarkable achievement, especially in light of decades worth of prior research on
the Prisoner's Dilemma in both the two-player and evolutionary settings.
\textit{ZD} strategies comprise a variety of new ways to play the Iterated
Prisoner's Dilemma, and Akin's generalization of \textit{cooperative} \textit{ZD}
to \textit{good} strategies provides novel insight into how
cooperation between two rational players can be stabilized.  However, in an evolutionary setting,
amongst both \textit{ZD} and \textit{good} strategies, it is the \textit{generous}
ones that are most successful.

\vspace{1cm}
%\begin{materials}
\section*{Appendix}

\noindent \textbf{Notation:} For ease of analysis, the parameter $\chi$
we use throughout is the inverse of that used by Press \&
Dyson \cite{Press:2012fk}.  In addition, to avoid confusion with
$\delta$-neighborhoods, we use $\lambda$ in place of Akin's $\delta$ \cite{Akin}.

%\subsection*{Evolutionary simulations}
\subsection*{Evolutionary simulations} We simulated a well-mixed population in which selection follows an
``imitation'' process \cite{Sig2,Traulsen:2006zr}.  At each, discrete time step, a
pair of individuals $(X,Y)$ is chosen at random. $X$ switches its strategy to imitate $Y$
with probability $f_{x\to y}$, 
\[
f_{x\to y}=\frac{1}{1+\exp[\sigma(s_x-s_y)]},
\]
where $s_x$ and $s_y$ denote the average IPD payoffs of players $X$ and $Y$ against
the entire population, and $\sigma$ denotes the strength of selection.
When a mutant strategy $X$ is introduced to a population otherwise consisting of 
a resident strategy $Y$, its probability of fixation, $\rho$ is given by
\begin{align}
\rho=
\frac{1}{\sum_{i=0}^{N-1}\prod_{j=1}^ie^{\sigma\left[(j-1)s_{yy}+(N-j)s_{yx}-js_{xy}-(N-j-1)s_{xx}\right]}}.
\end{align}
Taylor expansion to first order about $\sigma=0$ gives Eq.~1, the condition
for selective replacement of $Y$ by $X$ under weak selection.
%\subsection*{Evolution within the set of ZD strategies}

\subsection*{Evolutionary robustness of cooperative ZD strategies}
Suppose that a resident strategy $Y$ is
cooperative and ZD. We will show that $Y$ is evolutionary robust if and only if $\chi \geq (N+1)/(2N-1)$.
From Eq.~1 we deduce that 
$Y$ is robust against any mutant IPD strategy $X$ iff their payoffs satisfy
\[
s_{xy}(2N-1) \leq s_{yx}(N+1)+(B-C)(N-2).
\]
Using Eq.~2 to substitute for $s_{yx}$ yields the equivalent condition
\[
[(B-C)-s_{xy}](\chi(2N-1)-(N+1)) \geq 0.
\]
Furthermore, we know that $B-C \geq s_{xy}$ for any mutant $X$ (because, otherwise
Eq.~2 would imply that both $s_{yx}$ and $s_{xy}$ exceed $B-C$, which contradicts
the assumption $2R>T+S$). Therefore the cooperative ZD strategy $Y$ is 
robust if and only if $\chi \geq (N+1)/(2N-1)$.

\subsection*{Non-cooperative ZD strategies can be selectively replaced}
Here we show that a resident ZD strategy $Y$ with $\kappa_y<B-C$ is selectively replaced
by a ZD strategy $X$ with $\kappa_x=B-C$ and $\chi_x>(N+1)/(2N-1)$.
Since both players are ZD, their payoffs satisfy the equations
\begin{eqnarray}
\nonumber \chi_xs_{xy}-\chi_x\kappa_x&=&s_{yx}-\kappa_x\\
\nonumber \chi_ys_{yx}-\chi_y\kappa_y&=&s_{xy}-\kappa_y.
\end{eqnarray}
which result in the payoff matrix
\vspace*{-.1in}
\begin{table}[h] 
 %\caption{}
 \begin{center} \begin{tabular}{@{\vrule height 10.5pt depth4pt  width0pt}  p{0.25cm} | p{3.5cm} p{3.5cm}}\ \
 &\cellcolor[rgb]{1,1,1}{$X$} &\cellcolor[rgb]{1,1,1}{$Y$} \\ \hline
\cellcolor[rgb]{1,1,1}{$X$} &\cellcolor[rgb]{1.0,1.0,1.0}{$\kappa_x$} &\cellcolor[rgb]{1.0,1.0,1.0}{$\frac{(1-\chi_y)\kappa_y+\chi_y(1-\chi_x)\kappa_x}{ (1-\chi_y\chi_x)}$} \\
\cellcolor[rgb]{1,1,1}{$Y$} &\cellcolor[rgb]{1.0,1.0,1.0}{$\frac{(1-\chi_x)\kappa_x+\chi_x(1-\chi_y)\kappa_y}{ (1-\chi_y\chi_x)}$} &\cellcolor[rgb]{1.0,1.0,1.0}{$\kappa_y$} \\
\end{tabular} \end{center}
\end{table}
\noindent 
Substituting these payoffs into Eq.~1 shows that $X$ can selectively replace $Y$
iff
\begin{align}
\label{iniq}
&\left[(2(1-\chi_x)(1+\chi_y-N\chi_y)+(1-\chi_y)(1+\chi_x-N\chi_x)\right]\times&\\
\nonumber &(\kappa_x-\kappa_y)<0
\end{align}
%If we suppose $X$ is an extortioner ($\kappa_x=0$ and $\chi_x>0$), and $Y$ is generous ($\kappa_y=B-C$ and $\chi_y>0$) we find that selection favors $X$ replace $Y$ if
By our assumptions on $X$ and $Y$, $\kappa_x-\kappa_y > 0$ and $\chi_x>1/(N-1)$.
If $\chi_y>1/(N-1)$ then inequality (\ref{iniq}) is satisfied, 
and so $X$ is selected to replace $Y$.
If $\chi_y<1/(N-1)$ then, to determine the conditions for which $X$ is selected to replace $Y$
we make the coordinate transformation $\chi_{y}^*=\chi_{y}-\frac{1}{N-1}$ and $\chi_{x}^*=\chi_{x}-\frac{1}{N-1}$, 
so that $\chi_x^*>0$ and $\chi_{y}^*<0$.
The inequality (\ref{iniq}) is then satisfied provided
\[
-2\left(1-\chi_x^*-\frac{1}{N-1}\right)\chi_y^*-\left(1-\chi_y^*-\frac{1}{N-1}\right)\chi_x^*<0.
\]
Rearranging this gives
\[
\chi_x^*>\frac{-2\left(1-\frac{1}{N-1}\right)\chi_y^*}{\left(1-3\chi_y^*-\frac{1}{N-1}\right)}
\]
which is hardest to satisfy when $\chi_{y}^*$ is at its minimum, i.e.~$\chi_{y}^*=-1-\frac{1}{N-1}$.
This results in the inequality
\begin{equation}
\label{iniq2}
\chi_{x}>\frac{N+1}{2N-1}
\end{equation}
as a sufficient condition for $X$ to selectively replace $Y$. This sufficient condition is met
by our assumption on $X$.
And so non-cooperative ZD strategies can always be selectively replaced, provided $N>2$.

%As a result, we expect that evolution within ZD will tends towards 
%towards generosity, as exhibited in Fig.~1.

\subsection*{Generous strategies limit the difference between their payoff and their opponent's payoff}
Consider Eq.~2 for a
generous ZD strategy $Y$ facing an arbitrary opponent $X$:
\[
\phi\left[s_{xy}-\chi s_{yx}-(1-\chi)(B-C)\right]=0.
\]
Rearranging this expression gives the difference in the players' payoffs
\[
s_{xy}-s_{yx}=(1-\chi)((B-C)-s_{yx}).
\]
Increasing $\chi$ reduces the difference between two players' payoffs, 
regardless of the opponent's strategy. This is also true for generous good strategies, 
which satisfy
\[
s_{xy}-s_{yx}=(1-\chi)((B-C)-s_{yx})-\lambda(v_{cd}+v_{dc})
\]
where $v_{cd}$ denotes the equilibrium rate of the play $(cd)$ and $v_{dc}$ the
equilibrium rate of the play $(dc)$ \cite{Akin}.  The ability of generous strategies to limit
the difference in payoffs with arbitrary opponents accounts for their remarkable consistency
as invaders, as exemplified in Fig.~S8.  
On the other hand, a non-generous strategy, such as win-stay-lose-shift, 
is subject to larger differences between her payoff and her opponent's payoff,
leading to less consistent success as an invader (Fig.~S8).

\subsection*{Long-memory strategies}
%\noindent \textbf{Long-memory strategies} 
Our results for the evolutionary success
of generous strategies in a finite population also hold against longer-memory
opponents.  As per Press \& Dyson, from the perspective of a memory-1 player, a
long-memory opponent is equivalent to a memory-1 opponent.  Thus the payoff
$s_{xy}$ can be determined by considering only the set of memory-1 strategies.
However the payoff a long-memory opponent receives against itself, $s_{xx}$, may
depend on its memory capacity.  Nonetheless, under the standard IPD assumption
$2R>T+S$ the highest total payoff for any pair of players in the IPD is $2R$; and
so $s_{xx}\leq R$.  This
condition on $s_{xx}$ is the only condition required to derive our results on
the robustness of ZD and good strategies (see Supporting Information), and so our
results continue to hold even against long-memory invaders.

%\bibliography{Thesis2}

\bibliographystyle{pnas.bst}

\end{document}